\documentclass{article}
\usepackage{authblk}
\usepackage[utf8]{inputenc}
\usepackage{amsmath}
\usepackage{graphicx}
\usepackage[colorlinks=true, allcolors=blue]{hyperref}
\usepackage[round]{natbib}
\usepackage{hyperref}
\title{Comparing intervention measures in a model of a disease outbreak on a university campus}
\author{Best, A.}
\author{Singh, P.}
\affil{School of Mathematics \& Statistics, University of Sheffield, Sheffield, S3 7RH, UK}

\date{}

\begin{document}
\maketitle

\section*{Abstract}
A number of theoretical models have been developed in recent years modelling epidemic spread in educational settings such as universities to help inform re-opening strategies during the Covid-19 pandemic. However, these studies have had differing conclusions as to the most effective non-pharmaceutical interventions. They also largely assumed permanent acquired immunity, meaning we have less understanding of how disease dynamics will play out when immunity wanes. Here we complement these studies by developing and analysing a stochastic simulation model of disease spread on a university campus where we allow immunity to wane, expoloring the effectiveness of different interventions. We find that the two most effective interventions to limit the severity of a disease outbreak are reducing extra-household mixing and surveillance testing backed-up by a moderate isolation period. We find that contact tracing only has a limited effect, while reducing class sizes only has much effect if extra-household mixing is already low. We identify a range of measures that can not only limit an outbreak but prevent it entirely, and also comment on the variation in measures of severity that emerge from our stochastic simulations. We hope that our model may help in designing effective strategies for universities in future disease outbreaks.

\section{Introduction}

Understanding the dynamics of infectious disease spread continues to be a major area of research. Modelling approaches have for nearly a century used the classic compartmental Susceptible-Infected-Recovered (SIR) framework \citep{kermack27} to explore the possible dynamics in a range of settings. In recent years, the Covid-19 pandemic led to a huge growth in the field, with models often playing key roles in management strategies decided by policy makers \citep[see][and articles in the same special issue]{brookspollock21b}. Commonly these studies went well beyond the classic compartment model, in particular incorporating spatial and social networks to account for how an emerging infectious disease would likely spread through the population
\citep{kucharski20, ferguson20, firth20, danon21}. Incorporating such structures within populations are important since it is known networks with more 'local' interactions tend to lead to a lower epidemiological $R_0$ \citep{keeling99,keeling00}, a common measure for determining the speed of spread. Not only is this important for considering more realistic predictions of epidemic time-courses, but it implies that non-pharmaceutical interventions that limit an individual's contacts are likely to play an important role in controlling an epidemic \citep[e.g.][]{kain21,wren21}.

The focus of a number of modelling studies during the Covid-19 pandemic has been on educational settings such as schools and university campuses \citep{bahl20,brook20,cashore20,lopman20,borowiak20,paltiel20,best21a,endo21,woodhouse22}.  These studies have taken a range of methodological approaches, from deterministic to stochastic models, purely theoretical or fitted to data, and with a focus on just one intervention measure or many. Their differing assumptions have led to varying conclusions about the most effective non-pharmaceutical interventions to protect against epidemics. Most commonly they show that a blended approach \citep{brook20,brookspollock21} is most effective, with asymptomatic/surveillance testing with fast results often identified as a key element \citep{bahl20,brook20,paltiel20,brookspollock21,woodhouse22}. However, evidence for the importance of mixing restrictions, especially in terms of class sizes, appears mixed \citep{brookspollock21,endo21}.

While these recent studies have given us a good level of understanding of the spread of an emerging infectious disease on a university campus, open questions remain. In particular, the majority of these studies assumed that infection-acquired immunity was permanent, leaving an open question as to how the dynamics will differ when immunity wanes, where endemic disease will be expected in the long-run \citep{crellen21}. Here we use a stochastic simulation algorithm to look at a range of intervention measures against the spread of a novel infectious disease. Our focus is on the peak and total infected numbers during a 6 month period as well as estimates for the epidemiological $R_0$. By using a stochastic model we are able to examine the variation in possible outcomes as well as headline averages. We construct the model to be loosely representative of university campuses by assuming household structures and daily classes.

\section{Methods}

We developed a direct-method stochastic simulation algorithm \citep{gillespie77} in Python to explore an epidemic model, building on an earlier study \citep{best21a}. Python code for the model can be accessed at \url{https://github.com/abestshef/campus_epidemic}. The underlying epidemiological dynamics are SEIRS (Susceptible - Exposed - Infected - Recovered - Susceptible) with no births, deaths or migration. A population of 1000 individuals are initialised, with 980 initially susceptible and 20 infected, modelling the start of a novel epidemic. Infection requires direct contact between an infected and susceptible individual. If infection takes place, the individual initially becomes exposed, such that they cannot infect others (but would show up as a positive case when tested - see below). After a latent period with average 7 days, they go on to be fully infected. After an average of 7 further days they clear the disease and become recovered and immune. Immunity is not long-lasting, however, and after an average of 120 days recovered individuals return to being susceptible. Note that with the presence of waning immunity we would expect the infection to ultimately reach an endemic equilibrium if $R_0>1$ rather than epidemic burn-out in the mean-field equivalent model.

The 1000 individuals are divided into 100 households of 10. The majority of mixing is between these 10 housemates, but we allow some degree of random extra-household mixing, with a default of 5\% of interactions being extra-household. Individuals are also allocated to a single class that they attend each day for 2 hours (with the same class make-up every day). As a default, there are 5 classes of 200 individuals.

We then explore the impact of various interventions. Firstly, extra-household mixing can be reduced (for example due to imposed restrictions). Secondly, the size of classes can be reduced. In addition, we also allow for weekly surveillance testing of all individuals. These tests are assumed to be 95\% accurate at detecting positive cases (exposed and infected) and 99\% accurate at determining negative cases (susceptible and recovered). Results are assumed to be immediate. All those cases identified as positive then isolate fully and perfectly from the rest of the population for some set number of days. We also allow for contact tracing of positive cases, with a uniform random number of contacts, up to a given maximum, from the identified individual's household and class also told to isolate for the full isolation period. Note that we do not include symptomatic testing in the model, only surveillance testing.

We run the model for 180 days and record the time course of infection. Our primary focus in our analysis is on the peak number of infections (usually during the initial epidemic wave) and the total number of infections over the time period. We also find an estimate for $R_0$ by recording the actual number of infections generated by the initial 20 cases. We explore the impact of the different intervention methods on these metrics, visualising the output with violinplots to highlight the distribution of outputs (along with boxplots, which additionally highlight the median and inter-quartile range). We also use recent methods to visualise the 'most central' time-courses under certain interventions \citep{juul20}.

\section{Results}

\subsection{Overview: comparing interventions}

We first compare the outcome when one intervention (or linked interventions) is introduced against a default of no interventions. We compare the following cases:

\begin{itemize}
\item Default - no interventions.
\item Class size - the class sizes are reduced from 200 to 10.
\item Mixing - extra-household mixing is reduced by 90\% from the default.
\item Test \& isolate - 90\% of the population take weekly tests, with a 5-day isolation period for positive cases and no contact tracing.
\item Test, trace \& isolate (TTI) - 90\% of the population take weekly tests, with a more stringent 10-day isolation period, and up to 5 contacts are traced.
\item Blended - A mixed intervention strategy, with class sizes reduced to 50, extra-household mixing halved, 50\% of the population testing with a 7-day isolation period and up to 3 contacts traced.
\end{itemize}

\begin{figure}
\centering
\includegraphics[width=0.9\textwidth]{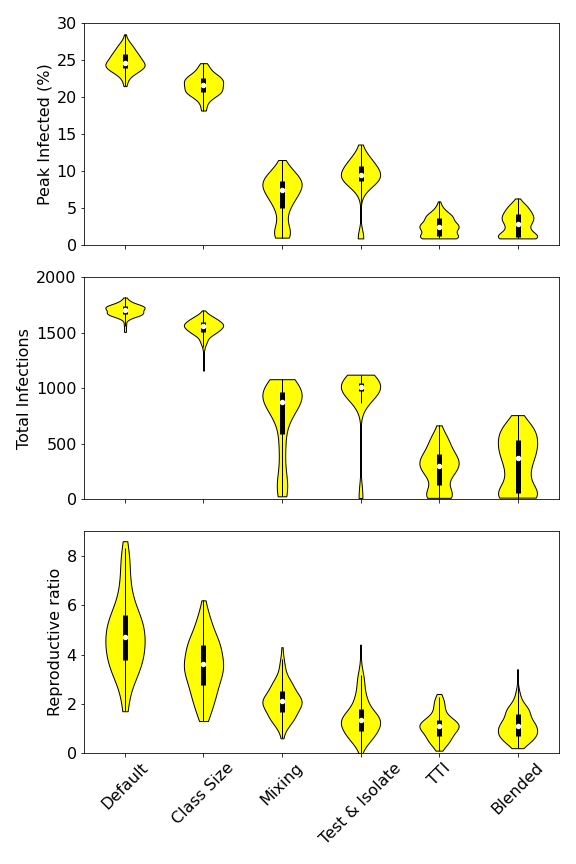}
\caption{Comparison of intervention measures on (top) peak infections (middle) total infections in 6 months and (bottom) estimated $R_0$. Default parameters: no mixing restrictions, class sizes 200, no testing.}
\end{figure}

\begin{table}
\begin{tabular}{ c | c | c | c}
Intervention & Median peak & Median total & Median $R_0$\\
  \hline			
  Default & 25\% & 1709 & 4.7 \\
  Class size & 22\% & 1557 & 3.6 \\
  Mixing & 7.5\% & 874 & 2.1 \\
  Weak TTI & 9.5\% & 1007 & 1.35\\
  Strong TTI & 2.5\% & 295 & 1.1\\
  Blended & 2.9\% & 367 & 1.1
\end{tabular}
\caption{Comparison of median measures of disease severity for different intervention strategies.}
\end{table}

The full distributions are shown in figure 1, with the medians also given in table 1. In the default case there is a large initial epidemic. It is noticeable that the high total infections means that many individuals have been infected (at least) twice - a clear consequence of our assumption of waning immunity. All of the single interventions have some effect on the severity of the outbreak, with the stronger measures often preventing an epidemic and/or leading to a disease-free state. Reducing the class size on its own has only a modest impact. Restricting extra-household mixing or the more limited testing and isolation strategy are more effective, reducing both the peak and total infections such that there is no overlap in their distributions with the default case. In fact, under both measures some outlier results occur where the disease dies out. The more stringent test, trace and isolate (TTI) intervention has a strong impact, with many simulations ending in the disease dying out. The distribution of peak and total infections for this stringent TTI regime is comparable with the blended approach with a range of intermediate interventions. In figure 2 we show the time-courses for the 50\% most central simulations for the default and blended interventions. This highlights the large effect the interventions have on limiting and even preventing the outbreak.

It is noticeable in figure 1 that the values for $R_0$ show considerable variation. In our default case, the values of $R_0$ vary from below 2 to more than 8, yet the distributions for the peak and total infections appear quite limited. Further, the distributions for $R_0$ under every intervention overlap with that of the default 'no intervention' case, yet the distributions for peak and total infected are very different. We believe this is driven by the fact that our estimates of $R_0$ are derived from the actual numbers of infections generated by the initial 20 cases. While realistic, this low base number likely leads to stochastic variation in how many cases are generated. 

\begin{figure}
\centering
\includegraphics[width=0.9\textwidth]{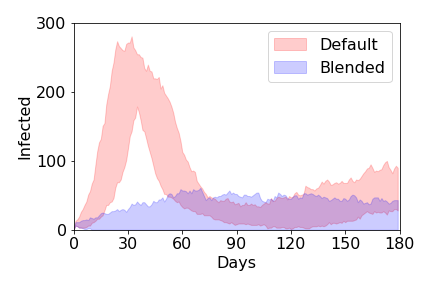}
\caption{Bounds of the 50\% 'most central' time-courses for infections numbers for the default (no interventions) and blended intervention strategy.}
\end{figure}

\subsection{Individual interventions}

We now look at each intervention individually in greater depth. In the following investigations we change our background case from 'no interventions' to 'minimal interventions' - random mixing is reduced by 20\%, classes are at 100 (default was 200), 20\% of the population test, and there is a 3 day isolation and a maximum of 1 individual contact-traced. This means we are assuming an immediate public health response of low interventions across the board, and can look at how different degrees of each intervention would impact the dynamics.

\subsubsection{Mixing restrictions}
Figure 3 demonstrates that mixing restrictions can have a significant impact on the epidemic. In particular, in the extreme case of no mixing outside of households and classes, for our parameters there are no significant outbreaks of disease in any of the simulations (maxmimum peak 1.5\% and maximum total 68). Even an 80\% reduction leads to many cases where no serious epidemic occurs (25th quartile peak 3.2\% and total 363), and a 60\% reduction has no overlap in the distributions for peak and total infections with the default 0\% reduction case.

\begin{figure}
\centering
\includegraphics[width=0.45\textwidth]{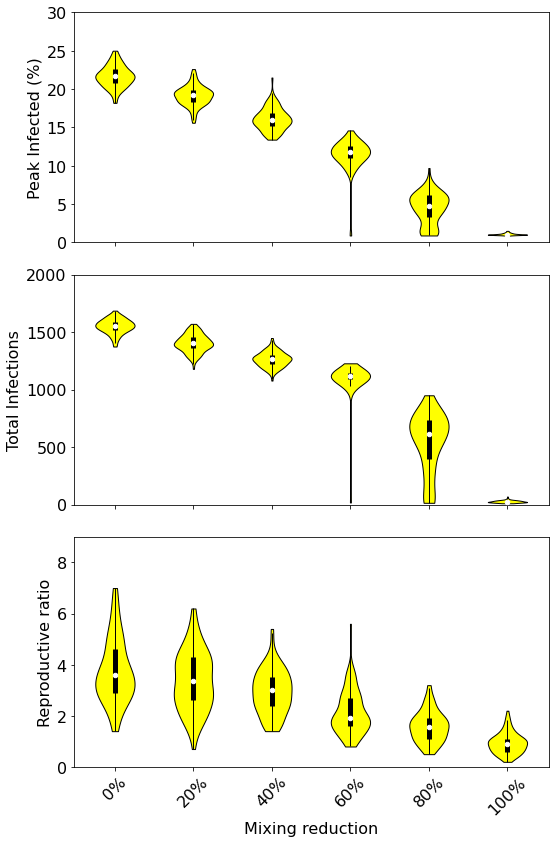}
\caption{Comparison of different reductions in extra-household mixing on (top) peak infections (middle) total infections in 6 months and (bottom) estimated $R_0$.}
\end{figure}

\subsubsection{Class sizes}

Reducing the class sizes appears to have only a modest impact on our measures of disease severity for the default of minimal interventions. Figure 4a shows that the distributions between the largest and smallest class sizes overlap for all measures. This is likely since in this case extra-household mixing remains at a reasonably high level (a 20\% decrease from the default), creating a more connected network that the class size limits do not overcome. Figure 4b shows the corresponding plots when mixing has been reduced by 80\%. In this case class sizes clearly do have an impact. Reducing the class size from 200 to 20 now reduces the median peak from 9.7\% to 1.6\% and the median total from 1029 to 178. Apart from a single outlier at the largest class size, the distributions for peak and total infections do not overlap with class sizes of 20 and smaller.

\begin{figure}
\centering
\includegraphics[width=0.45\textwidth]{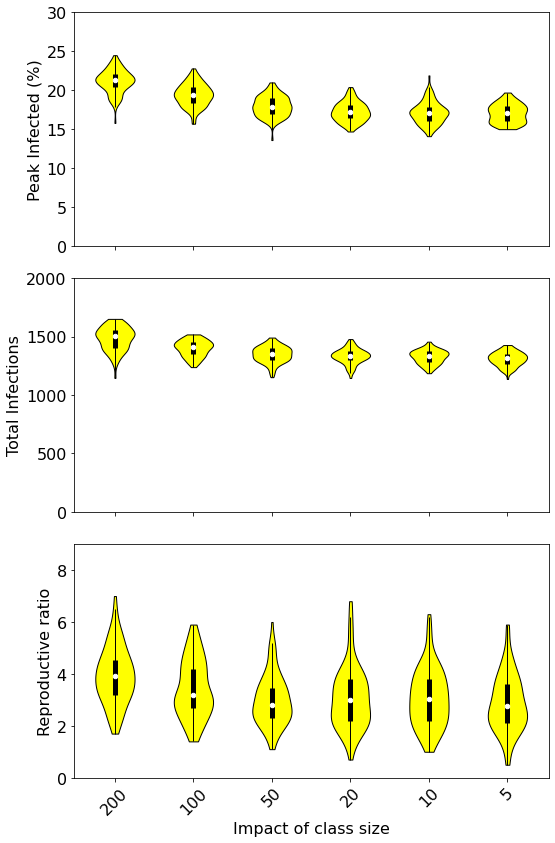}
\includegraphics[width=0.45\textwidth]{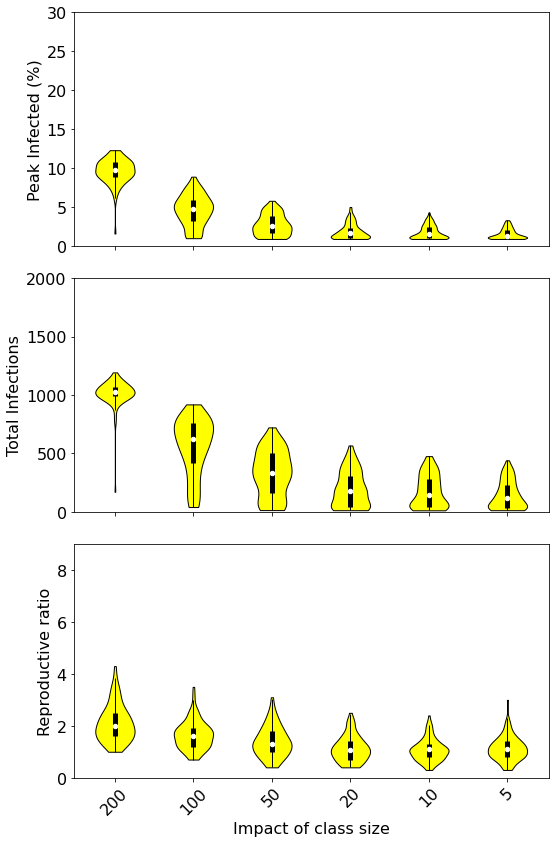}
\caption{Comparison of different reductions in class sizes on (top) peak infections (middle) total infections in 6 months and (bottom) estimated $R_0$.Left: mixing is reduced by 20\% from the default, right: mixing is reduced by 80\% from the default.}
\end{figure}

\subsubsection{Testing, tracing and isolation}

\begin{figure}
\centering
\includegraphics[width=0.45\textwidth]{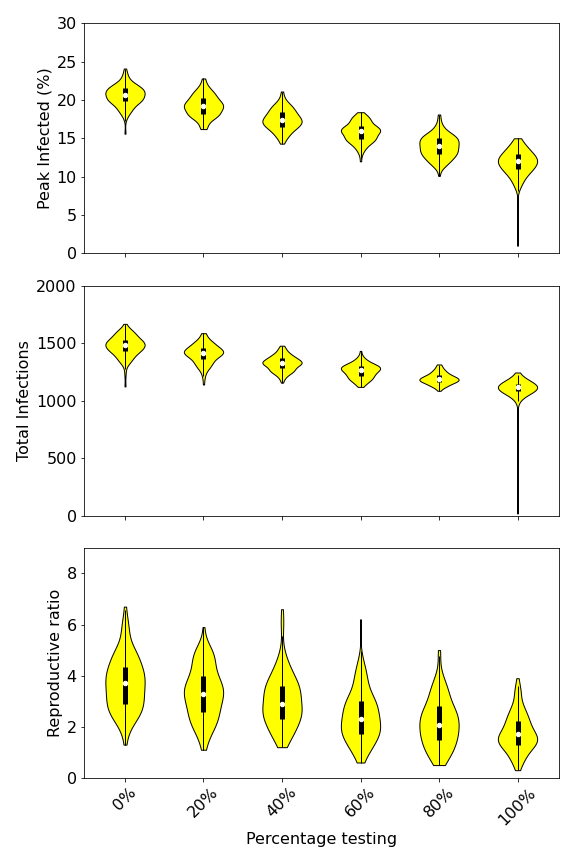}
\includegraphics[width=0.45\textwidth]{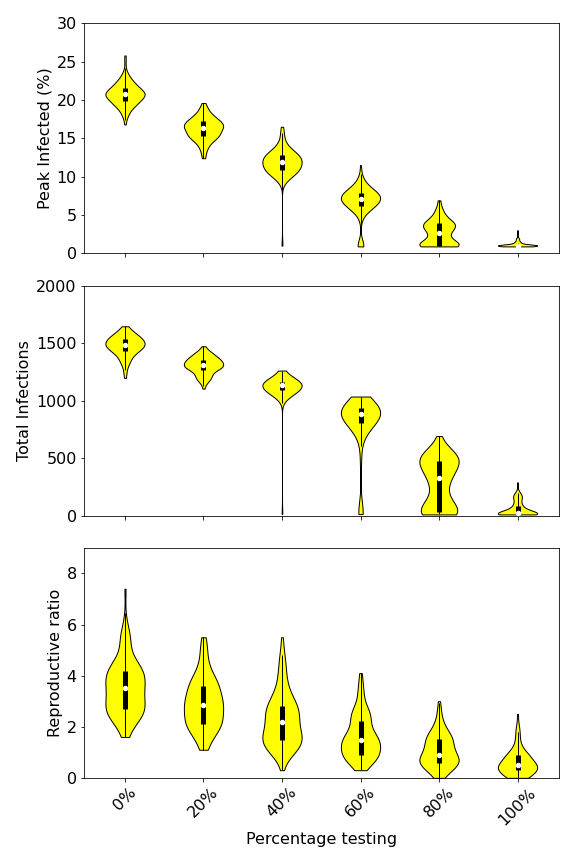}
\includegraphics[width=0.45\textwidth]{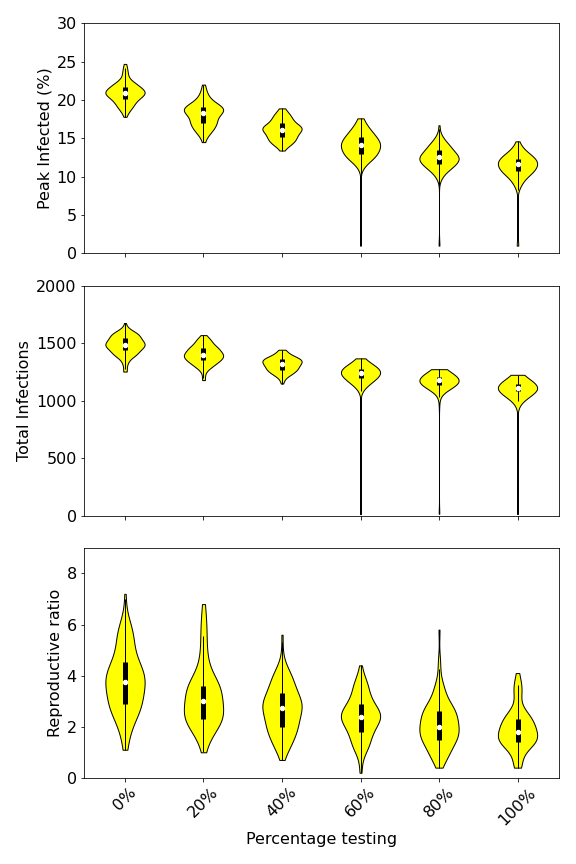}
\includegraphics[width=0.45\textwidth]{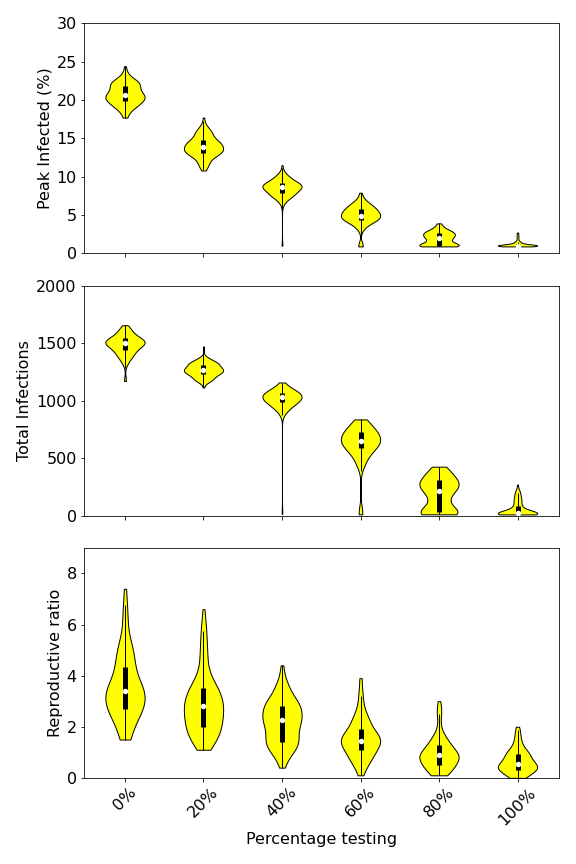}
\caption{Comparison of different proportions of weekly surveillance testing on (top) peak infections (middle) total infections in 6 months and (bottom) estimated $R_0$. Top-left: 3-day isolation and max. 1 contact traced; top-right: 10-day isolation and max. 1 contact traced; bottom-left:3-day isolation and max. 5 contacts traced; bottom-right: 10-day isolation and max. 5 contacts traced.}
\end{figure}

The interventions of testing, tracing and isolation must be considered as a package, since at the extreme if 100\% of the population test but there is no isolation period, the testing regime will have no effect. We first look at the default case of minimal isolation (3 days) and contact tracing (1 day) for varying proportions of the population testing, and then vary each of the secondary variables in isolation and together (figure 5).

In all cases, increasing the proportion of the population taking weekly tests decreases the peak and total infections and $R_0$. However, under the minimal tracing and isolation strategy this reduction is only slight (figure 5a). From 0\% to 100\% testing the median peak reduces from 20.7\% to 12.1\%, the median total from 1484 to 1117 and the median $R_0$ from 3.7 to 1.7, with only the distributions for the peak infected having no overlap between these extreme cases.

In contrast, when isolation is increased to 10 days a 100\% testing regime reliably prevents any significant outbreaks (maximum peak 2.8\%, maximum total 280), with the median $R_0<1$ (figure 5b). Even when 60\% test, there is no overlap in the distributions compared to when there is no testing, and in fact there is no overlap with the no testing case in the peak distributions when just 40\% test. It is worth recalling from figure 1 that even a 5 day isolation period can be effective when combined with high testing coverage.

Increasing the maximum number of contacts traced from 1 to 5 (with the isolation period returned to 3 days) has a small yet clear effect on reducing the epidemic (figure 5c). Now between the 0\% and 100\% testing extremes, the median peak reduces from 20.9\% to 11.7\% and the median total from 1488 to 1107. Comparing to the values from figure 5a, these reductions are small. However, now the peak distribution when testing is 60\% or higher does not overlap with the no testing case, but only the extreme 100\% testing distribution does not overlap with no testing for total infections. 

Finally, when both isolation and contact tracing are at their higher level, now only 80\% testing  is needed to reliably prevent significant outbreaks (maximum peak 3.9\%, maximum total 425), with the median $R_0<1$ (figure 5d). Even if 20\% test, the peak distribution shows no overlap with the no testing case. Comparing to figure 5b the increase in contacts traced makes a relatively modest difference to the distributions, whereas compared to figure 5c the increase in isolation period has made a significant difference.

In table 2 we summarise the median measures for a subset of the intervention combinations in figure 5, ranked in order of their effectiveness. These largely indicate that increasing testing is the most successful strategy, even with minimal isolation and contact tracing. Increasing isolation can also have a substantial impact, especially when combined with already high testing. For the parameters used here, increasing the number of contacts traced mostly has little effect. We can also notice that 20\% testing with a stronger package of isolation and contact tracing is very similar in its outcomes to 80\% testing with the minimal package of interventions.

\begin{table}
\begin{tabular}{ c | c | c | c | c | c}
Testing & Isolation & Contacts & Median peak & Median total & Median $R_0$\\
  \hline			
  20\% & 3 & 1 & 19.2\% & 1417 & 3.3 \\
  20\% & 3 & 5 & 18.4\% & 1395 & 3.0\\
  20\% & 10 & 1 & 16.4\% & 1308 & 2.9 \\
  20\% & 10 & 5 & 13.8\% & 1268 & 2.8 \\
  80\% & 3 & 1 & 14.0\% & 1187 & 2.1 \\
  80\% & 3 & 5 & 12.5\% & 1176 & 2.0\\
  80\% & 10 & 1 & 2.7\% & 330 & 0.9 \\
  80\% & 10 & 5 & 2.0\% & 216 & 0.6 
\end{tabular}
\caption{Comparison of median measures of disease severity for different combinations of testing, isolation and contact tracing.}
\end{table}

\section{Discussion}
We have used a stochastic simulation model to compare single and blended interventions to limit the spread of an epidemic in a closed population such as a university campus. Our results suggest that the two most effective intervention measures are (1) reduced extra-household mixing and (2) surveillance testing backed up by at least a moderate isolation period. Decreasing the class size only has a small effect under our model assumptions, as does increasing the number of contacts traced from each positively identified case. Overall, our model would conclude that a combination of moderate mixing restrictions and testing/isolation requirements would often be sufficient to not only limit the extent of an initial epidemic, but also prevent the disease persisting even when immunity is not long-lasting.  

Given that as a default extra-household mixing only accounts for 5\% of an individual's contacts outside of class in our model, it is notable how strong the impact of reducing mixing is. Picturing the network that is created, since mixing is random, even a low level can mean the whole population is still well connected, and an infectious disease can quickly spread. In contrast, under the more severe mixing restrictions the network quickly becomes constrained to the households and classes, and the speed of disease spread is substantially reduced. In contrast, the effectiveness of reducing class sizes depends on the level of extra-household mixing. This is because the make-up of classes is fixed creating a less connected population than from our assumption of random mixing. Previous modelling studies have considered mixing restrictions in a variety of ways, from limits on group sizes \citep{brook20,kain21}, to reducing the total size of the university population \citep{bahl20} to smaller class sizes \citep{best21a,brookspollock21}. These have all found that mixing restrictions reduce the severity of the epidemic but to different extents. The importance of smaller class sizes seems particularly variable between studies. In their data-driven model, \citet{brookspollock21} found that reducing class sizes was the single most effective intervention of the four tested, while the more general theoretical models of \citet{best21a} and \citet{borowiak20} also showed that small class sizes can significantly impact an epidemic when there is otherwise limited extra-household mixing. Yet in a data-driven study of influenza outbreaks in Japenese schools \citet{endo21} it was found that class sizes were only minimally associated with the rate of spread. Our work may give some context to these differing results as it suggests that the effectiveness of smaller class sizes depends on the underlying household and mixing assumptions. Overall, we expect that a mix of reduced social mixing and reduced class sizes would prove effective to prevent individuals from becoming too isolated while still maintaining some control over disease spread.

Testing, isolation and contact tracing must be considered as a package of strategies, since testing will have no impact if there is no isolation and vice versa. As we might expect, increasing all three interventions decreased the impact of the epidemic, but it is clear that an intermediate amount of surveillance testing backed by a moderate isolation period can be effective in limiting infection numbers, while a stronger package can reliably prevent an outbreak entirely. Interestingly, we found that the outcomes were comparable when 80\% of the population test but there are minimal isolation and contact tracing measures to when 20\% test but there are strong isolation and contact tracing measures. We found that increasing the number of contacts traced made more minor differences than increasing the proportion testing or lengthening the isolation period. This may be partly due to the fact that this figure is the maximum number traced, so the average number traced does not increase as substantially. However, it could also suggest that a basic contact tracing mechanism is simply not as effective at limiting future infections as other interventions considered. Previous studies similarly found that asymptomatic/surveillance testing of the population can be one of the most effective interventions \citep{bahl20,brook20,woodhouse22}, with \citet{brook20} further noting that such testing can reduce variation in the daily case count, making epidemics more predictable. Our results on testing agree with this, but with the exception that variation in numbers infected can increase close to the extinction boundary, partly due to our assumption of waning immunity making endemic disease the default outcome without interventions. Our model assumed that results were returned immediately, with a picture of the rapid lateral flow tests that have become common for Covid-19 testing in our minds. Previous studies that specifically included a delay in results found that reducing the wait time could lead to much reduced epidemics \citep{bahl20,brook20}, and as such we would expect the introduction of a delay in our model to lead to increases in infections.

Perhaps unsurprisingly, a blended package of the different interventions proves to be a good approach to prevent the severity of an epidemic and/or stop a disease becoming endemic, agreeing with previous modelling studies \citep{brook20,brookspollock21}. The precise restrictions needed will of course vary depending on the features of the infection. Under our modelling assumptions it would appear a rule of thumb to restrict or even prevent an outbreak would be for reducing extra-household mixing by around half alongside weekly surveillance testing of at least 50\% of the population with positive cases asked to isolate for 5-7 days. Reducing the class size and introducing contact tracing may also be used for more significant outbreaks. We found that the greatest variability in outcomes occurred with high levels of single interventions - for example reducing mixing by 80\% or testing 80\% of students. In these cases most simulations led to an intermediate severity of outbreak, but rare cases occurred where the epidemic never took off. It may be, therefore, that university campuses are fortunate in experiencing no epidemic with only limited controls. A recent study found that while most higher education institutions in the US enforced relatively low-cost non-pharmaceutical interventions against Covid-19 such as mandating mask wearing indoors, only 20\% of those sampled instituted all of the interventions recommended by the CDC including on-campus testing and restrictions on mixing \citep{moreland23}. Our work would suggest that, while undoubtedly costly and with important considerations for wellbeing, the more measures implemented the better for disease control. A number of theoretical studies have explored optimal control of an epidemic subject to constraints such as budget \citep{abakuks73,brandeau03,fan22}. Such methods could be applied to our model to ascertain the ideal blend of interventions under different budgets and targets.

We believe our model captures some key aspects of disease spread on a university campus - notably the household structures and class assignments - and may therefore provide a reasonable first estimation that could be of use when planning against future disease outbreaks. Our stochastic simulation approach provides an additional layer of detail in being able to describe the likelihood of different time-courses unlike a more simple, though tractable, deterministic model \citep{keeling99,sharkey08}. In particular it has highlighted how measures of $R_0$ from early case data may be very variable due to low numbers. In the extreme, estimates of $R_0=2$ were included in the distributions for both the worst-case 'no intervention' strategy (resulting in at least 1500 total infections and more than 20\% infected at the peak) and for the 'blended' strategy (resulting in fewer than 800 total infections and less than 7\% infected at the peak). As such, a very early estimate for $R_0$ may not be a highly reliable guide to the ultimate severity of the epidemic 

Of course there are many developments that may be explored for further realism. For example, we have assumed students belong to a single class, which we might expect to reduce infections compared to if students attend different classes each day. In contrast, we assumed extra-household mixing was entirely random, whereas the assumption of social groups might well reduce infections, and the use of real-world contact data as in previous studies \citep{firth20} could form the basis of such a model. We also assumed that individuals have no innate difference in their potential to infect, yet studies have shown that for many respiratory diseases certain individuals act as 'superspreaders' \citep{lloyd05}. However, it is unclear to what extent this is driven by innate infectiousness of the individual or their mixing patterns. The mixing patterns generated in our model will lead certain individuals to seed more infections than others, and it would be interesting to explore how different underlying mixing assumptions influence the emergence of superspreading. Including these or other potential additions will only add to the power of modelling approaches to predict and inform policy for disease outbreaks.

\bibliographystyle{spbasic}
\bibliography{biblio}

\end{document}